\documentclass{aa}
\usepackage{epsf}
\usepackage{natbib}
\bibpunct{(}{)}{;}{a}{}{,}

\newcommand{\Msun}{\ensuremath{M_{\odot}}}
\newcommand{\hi}{H~{\sc i}}
\newcommand{\himf}{H{\sc i}MF}
\newcommand{\kms}{\mbox{$\rm km\, s^{-1}$}}

\begin{document}
\title{An H{\sc i} survey of the Centaurus and Sculptor Groups}
\subtitle{Constraints on the space density of low mass galaxies}

\author{W.J.G. de Blok\inst{1}
\and M.A. Zwaan\inst{2}
\and M. Dijkstra\inst{3}
\and F.H. Briggs\inst{3}
\and K.C. Freeman\inst{4}}

\institute{Australia Telescope National Facility, PO Box 76, Epping, NSW 1710, Australia
\and School of Physics, Univ. of Melbourne, Parkville, VIC 3052, Australia
\and Kapteyn Astronomical Institute, PO Box 800, 9700 AV Groningen, The Netherlands
\and Research School of Astronomy \& Astrophysics, Mount Stromlo Observatory, Cotter Road, Weston ACT 2611, Australia}

\offprints{W.J.G. de Blok}

\date{Received ... / Accepted ...}

\abstract{We present results of two 21-cm \hi\ surveys
performed with the Australia Telescope Compact Array in the nearby
Centaurus A and Sculptor galaxy groups.  These surveys are sensitive
to compact \hi\ clouds and galaxies with \hi\ masses as low as $\sim 3
\times 10^6\ M_{\odot}$, and are therefore among the most sensitive
extragalactic \hi\ surveys to date.  The surveys consist of sparsely
spaced pointings that sample approximately 2\% of the groups' area on
the sky.  We detected previously known group members, but we found no
new \hi\ clouds or galaxies down to the sensitivity limit of the
surveys. If the HI mass function had a faint end slope of $\alpha =
1.5$ below $M_{HI} = 10^{7.5}\ M_{\odot}$ in these groups, we would
have expected $\sim 3$ new objects.  Cold dark matter theories of
galaxy formation predict the existence of a large number low mass DM
sub-halos that might appear as tiny satellites in galaxy groups.  Our
results support and extend similar conclusions derived from previous
\hi\ surveys that a \hi\ rich population of these satellites does not exist.
}

\titlerunning{HI survey of Cen and Scl groups}
\authorrunning{de Blok et al.}

\maketitle

\section{Introduction}

The Cold Dark Matter (CDM) theory is currently widely accepted as the
most successful theory for the formation of structure in the universe.
It has, however, become clear of late that two important results that
follow naturally from the CDM theory, are difficult to reconcile with
observational data: 1) The central parts of galaxy halos should
have high densities with cuspy distributions (e.g.\ \citealt{NFW96,
NFW97,moore99}) and, 2) high mass galaxies should be surrounded by large
numbers of compact, low-mass satellites (e.g.  \citealt{moore99b,
klypin99}).

With regard to the first problem, recent high-resolution rotation
curves of a large sample of Low Surface Brightness (LSB) and dwarf galaxy
rotation curves show that their dark matter distributions are best
characterized by a central constant density core with a typical size
of a few kpc \citep{dBMBR}. This is in sharp contrast with the CDM
result which predicts that the central parts of galaxies are
characterized by a central cusp with a density that increases sharply
towards the center.

Another problem is that the number of low-mass satellites actually
observed is much lower than predicted.  For example, in the Local
Group only $\sim 36$ dwarf galaxies are known, while CDM predicts
several hundreds of these low-mass galaxies (e.g.\
\citealt{moore99b,klypin99}).

If the gas densities at the centers of these objects are high enough
to prevent ionization by the metagalactic uv-background of a
significant fraction of the innate hydrogen gas, they should be
detectable at the present epoch in sensitive 21-cm \hi\ emission line
surveys.

This led to the hypothesis that the observed (compact) high velocity
clouds (HVCs) are actually extra-galactic and candidates for the
missing low mass halos \citep{blitz, brauner}.  Typical median
properties of the locally observed compact HVCs population as derived
from the HIPASS data \citep{putman} are a size of $\sim 0.2$ degrees,
a peak column density of $1.4 \times 10^{19}$ cm$^{-2}$, a velocity
width of 35 km s$^{-1}$ and a total flux density of 19.9 Jy km
s$^{-1}$. If compact HVCs are truly extra-galactic then this implies a
median \hi\ mass of $M_{HI}= 2.3 \times 10^{6}\ M_{\odot}$ for a
distance of $700$ kpc, or $M_{HI} = 7.5 \times 10^5\ M_{\odot}$ for a
distance of 400 kpc.  These \hi\ clouds should also be present in
other galaxy groups and, as the parameters above show, detectable.
There has recently been a large effort to find them in existing and
new \hi\ surveys
\citep{zwaanbriggs00,dahlem,zwaan01,verheijen00}. Groups are presently
the only environments where it is feasible to search for these clouds
as the overdensity of the groups increases the number of possible
detections.  Since in numerical simulations of hierarchical structure
formation the number of low mass halos is scale-invariant, the same
ratio of low mass halos to normal size galaxies is expected in groups
and in the field.  The surveys showed that HVC-like objects are
\emph{not} distributed throughout groups and galaxy halos on scales of
$\sim 1$ Mpc (but see \citealt{brbur}).  The HVCs must therefore be
much closer to our Milky Way Galaxy than \citet{blitz} originally
proposed and they are therefore insignificant contributors to the
baryon and total mass content of groups and galaxies.

There is, however, still a region of parameter space that has not been
surveyed extensively.  \hi\ surveys to date have been less suited to
constrain the population of {\em compact\/}, high column density, low
\hi\ mass clouds and galaxies.  If such a population were to exist, it
could contain a substantial fraction of $\Omega_{\rm gas}(z=0)$ locked
up in such low \hi\ masses which would not have been detected in recent \hi\
surveys \citep{rosenberg,zwaan97,kilborn}.

In this paper we present results of two surveys designed to probe the
regime of compact, low \hi\ mass objects.  These surveys, conducted
with the Australia Telescope Compact Array (ATCA), have the lowest
\hi\ mass sensitivity to date with high spatial and velocity
resolution.  Section 2 discussed the targets of our surveys; section~3
addresses the observations.  In section 4 we discuss the results, and
section 5 describes the implications for the space density of low-mass
galaxies.

\section{Targets}

At 3.5 Mpc distance from the Milky Way, the Cen A group is one of the
nearest gas-rich groups known.  It is well-defined both in projection
on the sky and in velocity
\citep{cote}.  It contains a large variety of galaxies, including the
well-known radio galaxy Cen A (NGC 5128).  The largest members are
generally early-type galaxies.  A large number of dwarf galaxies is
however known \citep{cote, cote00, banks}, making this group ideal for
this kind of survey, due to its small distance as well as large over
density.

The Sculptor group has different properties.  It is dominated by a
small number of gas-rich late-type spiral galaxies (e.g.
\citealt{jerjen, cote00}).  Sculptor is thought to be an elongated
collection of galaxies seen pole-on \citep{jerjen} and linked with the
Local Group.  Following \citet{cote} we adopt a distance of 2.5 Mpc.
A previous \hi\ survey of the Scl group \citep{haynes} detected many
\hi\ clouds, but these are not associated with the group. Rather they belong
to the Magellanic Stream and the Milky Way.

\citet{cote} derive from their extended Cen A and Scl dwarf samples
velocity dispersions of 150 and 202 \kms\ respectively. The equivalent
crossing times are $4.5 \times 10^9$ and $3.2 \times 10^9$ years. This
is in both cases a significant fraction of the Hubble time, showing
that both groups are dynamically unevolved, loose structures.

\section{Observations}

We have observed regions in both the nearby Centaurus A group and the
Sculptor group in \hi\ with the ATCA. As we have used different
observing strategies for each group we discuss them separately.

\subsection{Centaurus}
We observed part of the Cen A group at 21-cm using a grid of 36
pointings with the Australia Telescope Compact Array in the 750-D
configuration. Figure 1 overlays the pointing centers on the
distribution of member galaxies \citep{cote} on the sky.  The grid
includes M83, and several known dwarf galaxies, and a large
underdensity located behind M83.  We thus sample a variety of
densities.  Table~1 lists the coordinates of the pointing centres for
all pointings.  Pointings in the grid were spread 1 degree apart.

The grid was observed over 9 runs of 12 hours each (3-4 and 12-18 July
1999). In each 12 hour session one block of 4 pointing was observed by
cycling over the 4 pointings in mosaicing mode (90 seconds per
pointing per cycle). We used a bandwidth of 8 MHz divided into 1024
channels. The velocity range was 0 - 1700 km s$^{-1}$ giving a good
coverage of the velocity spread of the galaxies in the group.

For each pointing a data cube was created measuring $256 \times 256
\times 785$ pixels and a pixel size of $16'' \times 16'' \times 2$ km
s$^{-1}$, thus encompassing two primary beam FWHM per pointing.  The
average noise for all cubes was 10.3 mJy per channel (without Hanning
smoothing; no primary beam correction was applied).  Table 1 also
lists the average noise for each cube separately.  The minimum
detectable \hi\ mass per 2 km/s channel in a 3 hour integration is
thus $\sim 3\times 10^5 M_\odot$ ($5\sigma$) at the pointing center.
For a velocity width of 20 km s$^{-1}$, the minimum detected \hi\ mass
would be $\sim 3 \times 10^6$ $M_{\odot}$, assuming optimal velocity
smoothing.  The synthesized beam was $65''\times 48''$, corresponding
to $\sim 1 $ kpc at the distance of Cen A.  The column density sensitivity
($5\sigma$) for each 2 \kms\ channel is $4\times 10^{19}$ cm$^{-2}$ or
$8\times 10^{19}$ after smoothing to 10 \kms.

As the grid points are spread 1 degree apart, and the primary beam
FWHM of the Compact Array measures 35$'$ at 21-cm, this grid will not
give uniform sensitivity coverage, but that is specifically not the
intention of this part of the experiment. For any (moderately) steep
H{\sc i} Mass Function (H{\sc i}MF) the number of low mass galaxies in
a unit volume will be larger than the number of high mass galaxies. A
larger volume thus needs to be probed in order to get good statistics
on the high mass galaxies. However, high mass galaxies in general have
larger fluxes and can be found out to a larger distance from the
center of the primary beam than low-mass, low flux galaxies. By
spreading out the grid points we are therefore able to probe the
dwarfs in the central parts of the primary beam, while the outer parts
of the primary beam will give us enough volume coverage to pick up
sufficient high mass galaxies.  This survey probes approximately 4\%
of the group volume at $M_{\rm HI}=10^6 M_\odot$ and about 12\% at
$10^7 M_\odot$ (since this mass can be detected at larger distances
from the center of the primary beam). These numbers are based on a
group radius of 0.64 Mpc \citep{vandenbergh}

\begin{figure} 
\epsfxsize=\hsize 
\epsfbox[85 200 592 718]{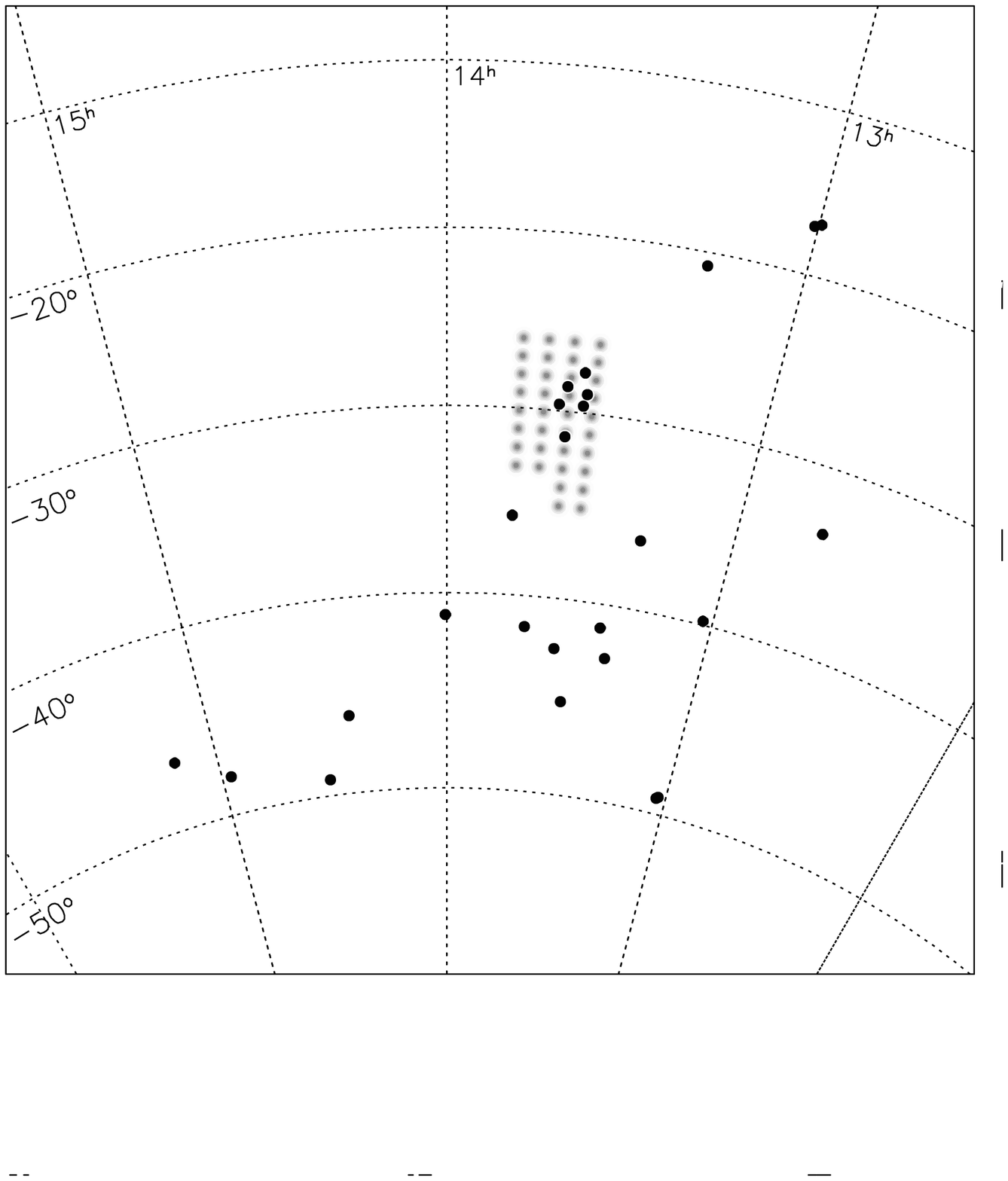}
\caption{The observed grid on the sky of the Cen survey. Observed
pointings are indicated by grey circles. Known galaxies from the
catalog by C\^ot\'e et al (1997) are indicated by black points.}
\end{figure}

This mode of observing probes a different area of parameter space than
blind extragalactic \hi\ surveys such as AHISS \citep{zwaan97},
HIPASS \citep{hipass} or ADBS \citep{rosenberg}. These surveys are
conducted with considerably coarser spatial and velocity resolution
(for HIPASS $15'$ and $13.2$ km s$^{-1}$ respectively), but with a
large area coverage.  They are excellent at picking up extended low
surface brightness emission, e.g.\ local HVCs, but their coarse
resolutions are insufficient to resolve e.g.\ small \hi\ clouds in
the close vicinity of other galaxies \citep{barnes}. Their relatively
coarse velocity resolutions furthermore make it difficult to
distinguish narrow-velocity width (and presumably low-mass) galaxies
from noise or interference spikes without further follow-up
observations.  Our current survey has much higher spatial and velocity
resolution, but covers a smaller area. It does however enable us to
look 10 times lower down the slope of the \himf .  A survey such as the
current one is better suited to pick up compact and narrow-line width
objects and is therefore complementary to the low resolution blind
\hi\ surveys.

\subsection{Sculptor}

Our original intention for the Scl survey was to image a large area
with uniform sensitivity, with a fully sampled grid of pointings, as
opposed to the Cen survey where the pointing were all independent.
The motivation for this was to search in the volume between Sculptor
and the Local Group. The Scl survey was done in the same observing
session as the Cen survey, at those times when Cen was below the
horizon.

We chose an area centered on $0^h30^m, -30^{\circ}00'$ which we
observed using a hexagonal grid with a 19.6$'$ distance between the
pointings. In each 12h session, we  observed 4 pointings, cycling
over them every 15 min. We used a 4 MHz bandwidth, centered on a
velocity of 500 km s$^{-1}$.  Unfortunately, as the Scl survey
happened in unallocated time, we were limited by telescope overrides
and maintenance. This meant that we could observe only 22 pointings in
a rather narrow strip and the area with uniform noise in the final
mosaic thus proved to be rather small. In addition, some instrumental
problems resulted in some pointings having a higher noise level. In
the end we decided that the only way to get reliable statistics was to
treat all pointings as independent pointings and proceed in the same
manner as with the Cen survey.  Table 1 lists the coordinates of the
22 pointings that we observed. Fig.~2 shows their lay-out on the sky.
Also shown are the positions of the Sculptor group members \citep{jerjen00}

\begin{figure}
\epsfxsize=\hsize
\epsfbox[85 200 592 718]{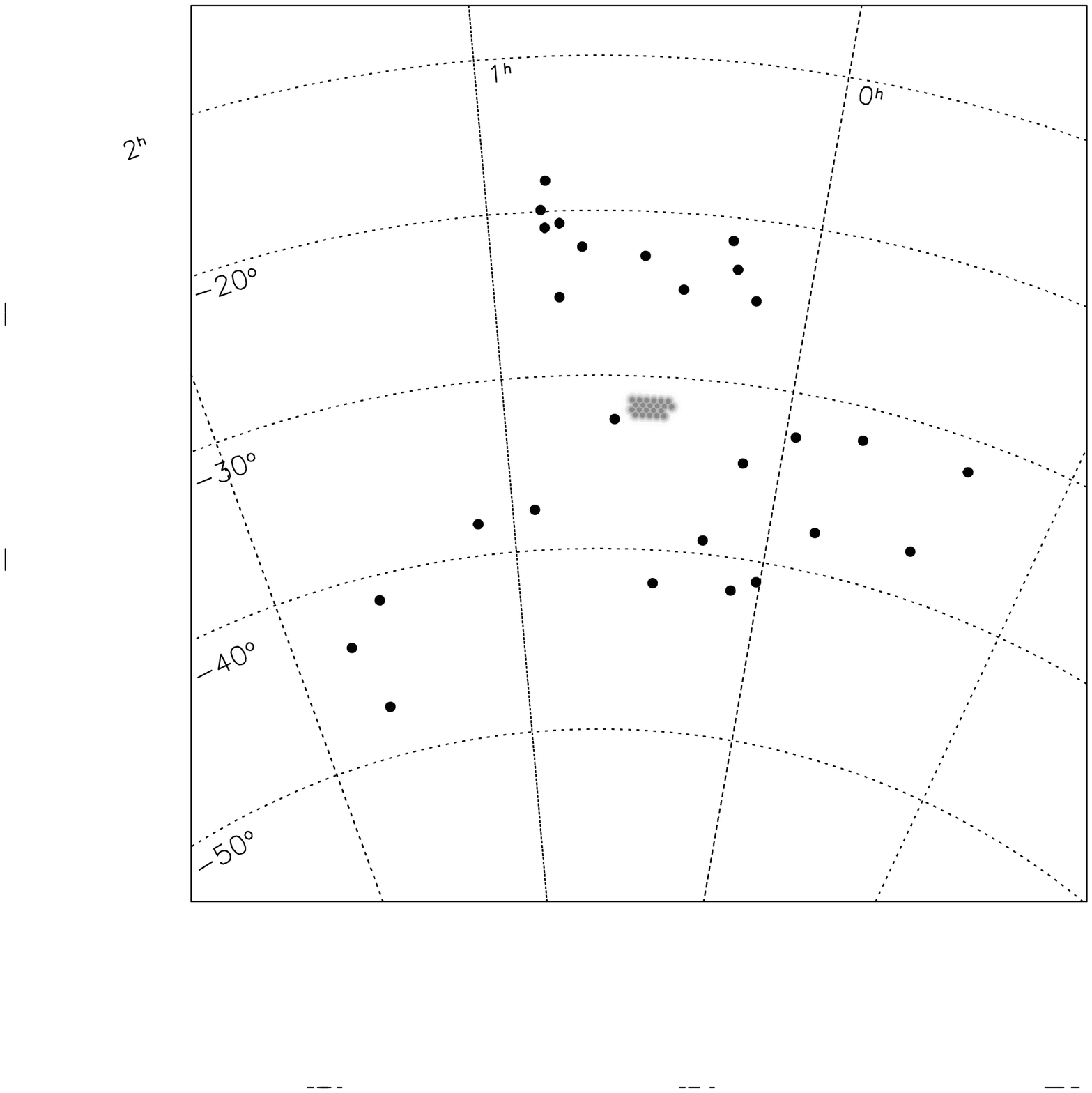}
\caption{The observed grid on the sky of the Scl survey. Observed
pointings are indicated by grey circles. Known galaxies are indicated
by black points.}
\end{figure}

The individual pointings were gridded in data cubes using the same
pixel size as the Cen cubes (i.e. $16'' \times 16'' \times 2$ km
s$^{-1}$). The final cubes measured $256 \times 256 \times 410$
pixels, with the velocity axis running from $-310$ to 530 km
s$^{-1}$. The synthesized beam size is $57'' \times 29''$. The average
noise per 2 km s$^{-1}$ channel was 10.2 (15.5) mJy for pointings
A1-D4 (E1-F2).  The column density sensitivity ($5\sigma$) for each 2
\kms\ channel is $7 (10) \times 10^{19}$ cm$^{-2}$ or $1.5 (2.3)
\times 10^{20}$ after smoothing to 10 \kms.

\section{Analysis}

\subsection{Centaurus}

\subsubsection{Known Objects}

In a number of fields previously catalogued objects were present
within a $30'$ radius of the field center and within the $0-1700$ km
s$^{-1}$ velocity range. These fields are indicated with an asterisk
in Table 1. These objects are listed in Table 2 along with their parameters
as derived from the HIPASS data base.

A fair fraction could however not be retrieved from the ATCA data.
These were usually objects that were far from the field center, where
primary beam attenuation makes their signal too weak to retrieve.  In
total, four objects were detected. We list their parameters as derived
from the ATCA data in Table 2 as well.  Figure 3 shows integrated
column density maps and spectra of three of the four detections (we do
not show the M83 data: these are severely affected by missing spacing
problems).

The two galaxies with \hi\ masses of a few times $10^8$ \Msun\ are
easily detected.  The observation of IC 4316 with an \hi\ mass of
$3\cdot 10^7$ \Msun\ is almost at the limit of what the eye can see in
the data. IC 4316 is $\sim 17'$ from the pointing center and primary
beam attenuation has reduced the signal to $\sim 0.55$ of the value it
would have in the pointing center.  As the velocity width of IC 4316
is $\sim 40$ km s$^{-1}$, it follows that the minimum \hi\ mass that
we could detect at the pointing center would be $\sim 8
\times 10^{6}$ \Msun\ for a 20 \kms-wide profile.
Optimal smoothing would improve this limit by a factor of a few,
bringing us close to the theoretical limit of the data.  We will
return to this after discussing the blind search.

\begin{table}
\caption{Positions Field Centers}
\begin{tabular}{lllr}
\hline
Name & $\alpha (2000.0)$ & $\delta (2000.0)$ &$\sigma$ (mJy) \\
\hline
\multicolumn{4}{c}{Centaurus}\\
\hline
A1&	13:32:24.00&	-27:51:30.00&10.6\\
A2&	13:37:00.00&	-27:51:30.00&10.4\rlap{ *}\\
A3&	13:37:00.00&	-28:51:30.00&10.5\rlap{ *}\\
A4&	13:32:24.00&	-28:51:30.00&10.3\\

B1&	13:41:36.00&	-27:51:30.00&10.3\\
B2&	13:46:12.00&	-27:51:30.00&10.0\\
B3&	13:46:12.00&	-28:51:30.00&10.3\\
B4&	13:41:36.00&	-28:51:30.00&10.0\rlap{ *}\\

C1&	13:32:24.00&	-29:51:30.00&11.0\\
C2&	13:37:00.00& 	-29:51:30.00&10.7\rlap{ *}\\
C3&	13:37:00.00&	-30:51:30.00&11.1\rlap{ *}\\
C4&	13:32:24.00&	-30:51:30.00&10.6\\

D1&	13:41:36.00&	-29:51:30.00&10.2\rlap{ *}\\
D2&	13:46:12.00&	-29:51:30.00&9.8\\
D3&	13:46:12.00&	-30:51:30.00&10.1\\
D4&	13:41:36.00&	-30:51:30.00&9.8\\

E1&	13:32:24.00&	-31:51:30.00&10.3\\
E2&	13:37:00.00&	-31:51:30.00&10.0\\
E3&	13:37:00.00&	-32:51:30.00&10.3\\
E4&	13:32:24.00&	-32:51:30.00&10.1\\

F1&	13:41:36.00&	-31:51:30.00&10.2\rlap{ *}\\
F2&	13:46:12.00&	-31:51:30.00&10.1\\
F3&	13:46:12.00&	-32:51:30.00&10.3\\
F4&	13:41:36.00&	-32:51:30.00&10.1\\

G1&	13:32:24.00&	-33:51:30.00&11.5\\
G2&	13:37:00.00&	-33:51:30.00&11.2\\
G3&	13:37:00.00&	-34:51:30.00&11.1\\
G4&	13:32:24.00& 	-34:51:30.00&10.9\\

H1&	13:41:36.00&	-33:51:30.00&10.4\\
H2&	13:46:12.00&	-33:51:30.00&10.0\\
H3&	13:46:12.00&	-34:51:30.00&10.2\\
H4&	13:41:36.00&	-34:51:30.00&10.0\\

I1&	13:32:24.00&	-35:51:30.00&9.9\\
I2&	13:37:00.00&	-35:51:30.00&9.5\\
I3&	13:37:00.00&	-36:51:30.00&9.7\\
I4&	13:32:24.00&	-36:51:30.00&9.6\\
\hline
\multicolumn{4}{c}{Sculptor}\\
\hline
A1 & 0:33:00.24 & -31:25:58.80 & 12.5\\
A2 & 0:32:15.12 & -31:43:01.20 & 12.7\\
A3 & 0:30:44.88 & -31:43:01.20 & 12.7\\
A4 & 0:31:30.00 & -31:25:58.80 & 13.0\\
B1 & 0:30:00.00 & -31:25:58.80 & 13.9\\
B2 & 0:29:14.88 & -31:43:01.20 & 14.3\\
B3 & 0:27:44.64 & -31:43:01.20 & 14.2\\
B4 & 0:28:29.76 & -31:25:58.80 & 14.5\\
C1 & 0:26:59.76 & -31:25:58.80 & 10.5\\
C2 & 0:26:14.64 & -31:43:01.20 & 10.6\\
C3 & 0:24:44.40 & -31:43:01.20 & 10.6\\
C4 & 0:25:29.52 & -31:25:58.80 & 10.6\\
D1 & 0:33:00.24 & -32:00:00.00 & 10.5\\
D2 & 0:32:15.12 & -32:16:58.80 & 10.5\\
D3 & 0:30:44.88 & -32:16:58.80 & 10.6\\
D4 & 0:31:30.00 & -32:00:00.00 & 10.7\\
E1 & 0:30:00.00 & -32:00:00.00 & 15.3\\
E2 & 0:29:14.88 & -32:16:58.80 & 15.3\\
E3 & 0:27:44.64 & -32:16:58.80 & 15.4\\
E4 & 0:28:29.76 & -32:00:00.00 & 15.8\\
F1 & 0:26:59.76 & -32:16:58.80 & 15.7\\
F2 & 0:26:14.64 & -32:00:00.00 & 15.5\\
\hline
\end{tabular}
\end{table}

\begin{table*}
\caption{Known galaxies in survey volume}
\begin{tabular}{lllllrlcrrrl}
\hline
Field &	Name  &	$\alpha$ (2000.0) &$\delta$ (2000.0)&$V_{hel}$& $S$ & ${W_{20}}$ & $\log M_{HI}$&$S$ & ${W_{20}}$&$R$ &Notes\\
  &              &            &             &    &\multicolumn{3}{c}{ATCA}&\multicolumn{2}{c}{HIPASS}&&\\
\hline
(1)&(2)&(3)&(4)&(5)&(6)&(7)&(8)&(9)&(10)&(11)&(12)\\
\hline
A2&ESO 444- G 084& 13h37m20.1s& --28d02m46s&  583 & 210.3& 75 & 8.78 & 223.8 & 75 & 12.1  &  \\
A3&UGCA 365      & 13h36m30.7s& --29d14m11s&  577 &  --  & -- & --   & 177.0 & 52 & 23.6 & \\
B4&IC 4316	 & 13h40m18.1s& --28d53m41s&  577 & 11.5 & 43 &7.51&$\sim 16.0$&40& 17.2 &  \\
C2&M83	         & 13h37m00.8s& --29d51m59s& (516)& \multicolumn{3}{c}{detected}& \multicolumn{2}{c}{detected}&0 &\\
C2&[KK98] 208    & 13h36m35.5s& --29d34m17s& (381)&  --  & -- & --   & -- & -- & 18.0    & c \\
C2&MCG -05-32-042& 13h35m07.9s& --30d07m05s& (300)&  --  & -- & --   & -- & -- & 28.9    &b\\
C3&ESO 444- G 082& 13h37m07.3s& --30d58m59s& (526)&  --  & -- & --   & -- & -- & 7.6    & b \\
D1&NGC 5264      & 13h41m36.9s& --29d54m50s&  482 & 152.1& 44 & 8.64 & 265.1 & 52 &3.3  &     \\
F1&ESO 445- G 007& 13h40m20.9s& --31d42m04s& 1659 &  --  & -- & --   & 315.6 & 100& 9.0 &    a\\
F1&NGC 5253      & 13h39m55.9s& --31d38m24s&  407 &  --  & -- & --   & 522.4 & 104& 25.0  &     \\
\hline
\end{tabular}
\begin{minipage}{16.8cm}
{\bf Notes}: (1) Field ID. (2) Galaxy identification. (3) Right
Ascension (2000.0). (4) Declination (2000.0). (5) Heliocentric
velocity. ATCA used where available, else HIPASS value. Value between
brackets denotes catalog value. (6) Total flux ATCA observation in Jy
km s$^{-1}$. (7) ATCA velocity width at 20\% level. (8) Logarithmic
ATCA \hi\ mass. (9) Total flux HIPASS observation in Jy km
s$^{-1}$. (10) HIPASS velocity width at 20\% level. (11) Distance in arcmin to nearest field center.
(12) Additional
notes: (a) not detected: edge of band (b) not detected in HIPASS (c) confused with
M83 in HIPASS
\end{minipage}
\end{table*}

\begin{figure*}
\epsfxsize=0.33\hsize
\epsfbox[74 54 447 356]{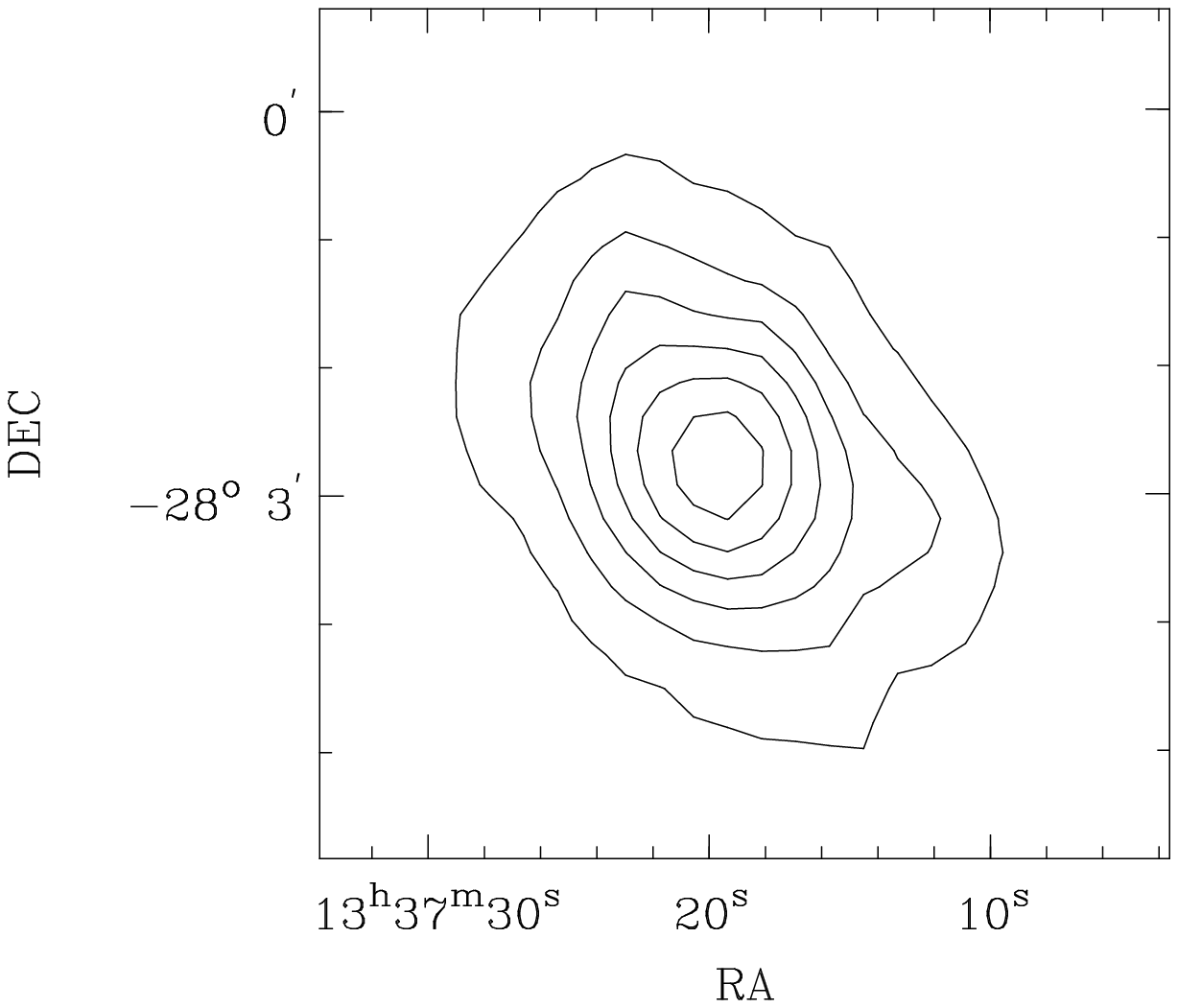}
\epsfxsize=0.33\hsize
\epsfbox[74 54 447 356]{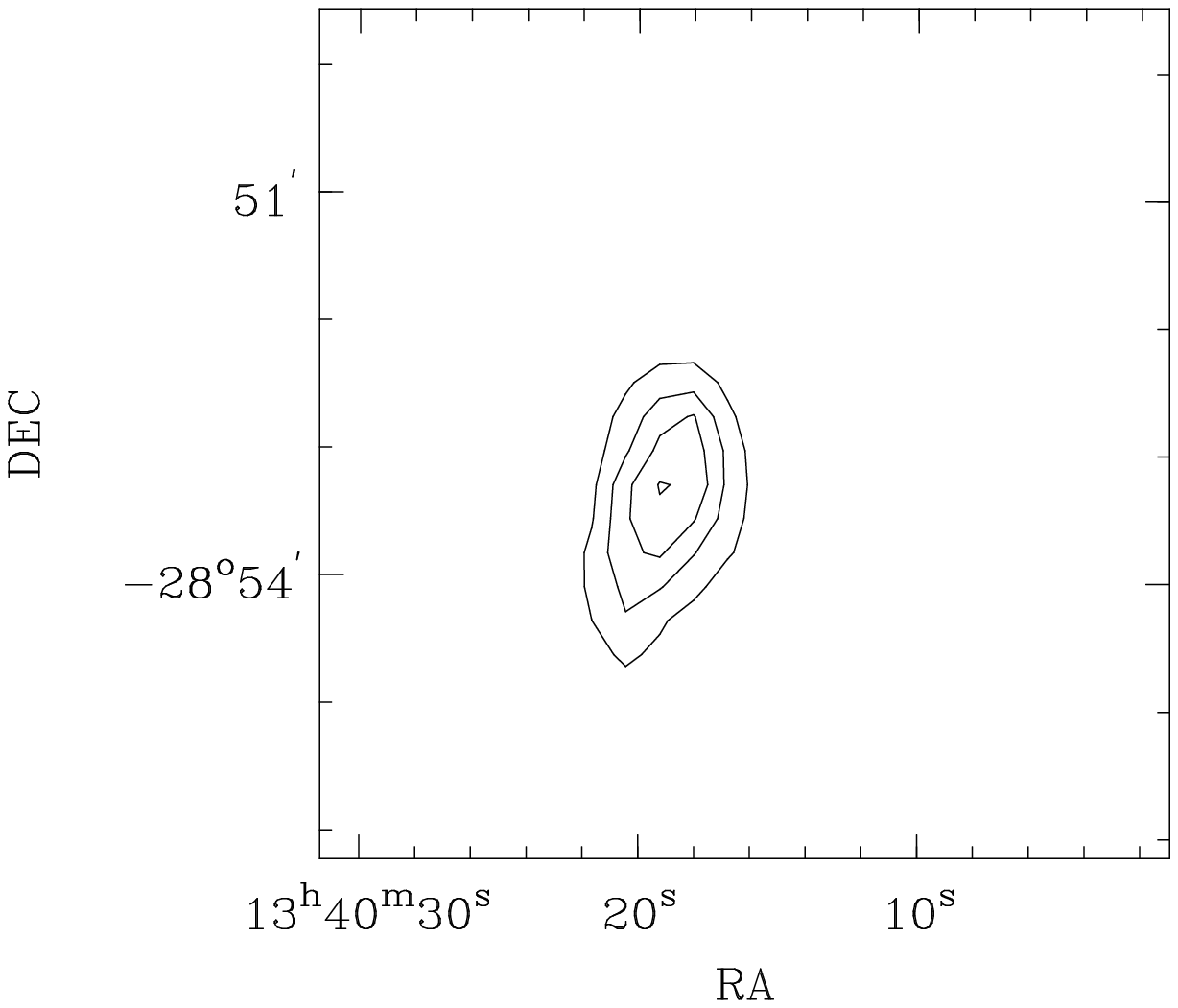}
\epsfxsize=0.33\hsize
\epsfbox[74 54 447 356]{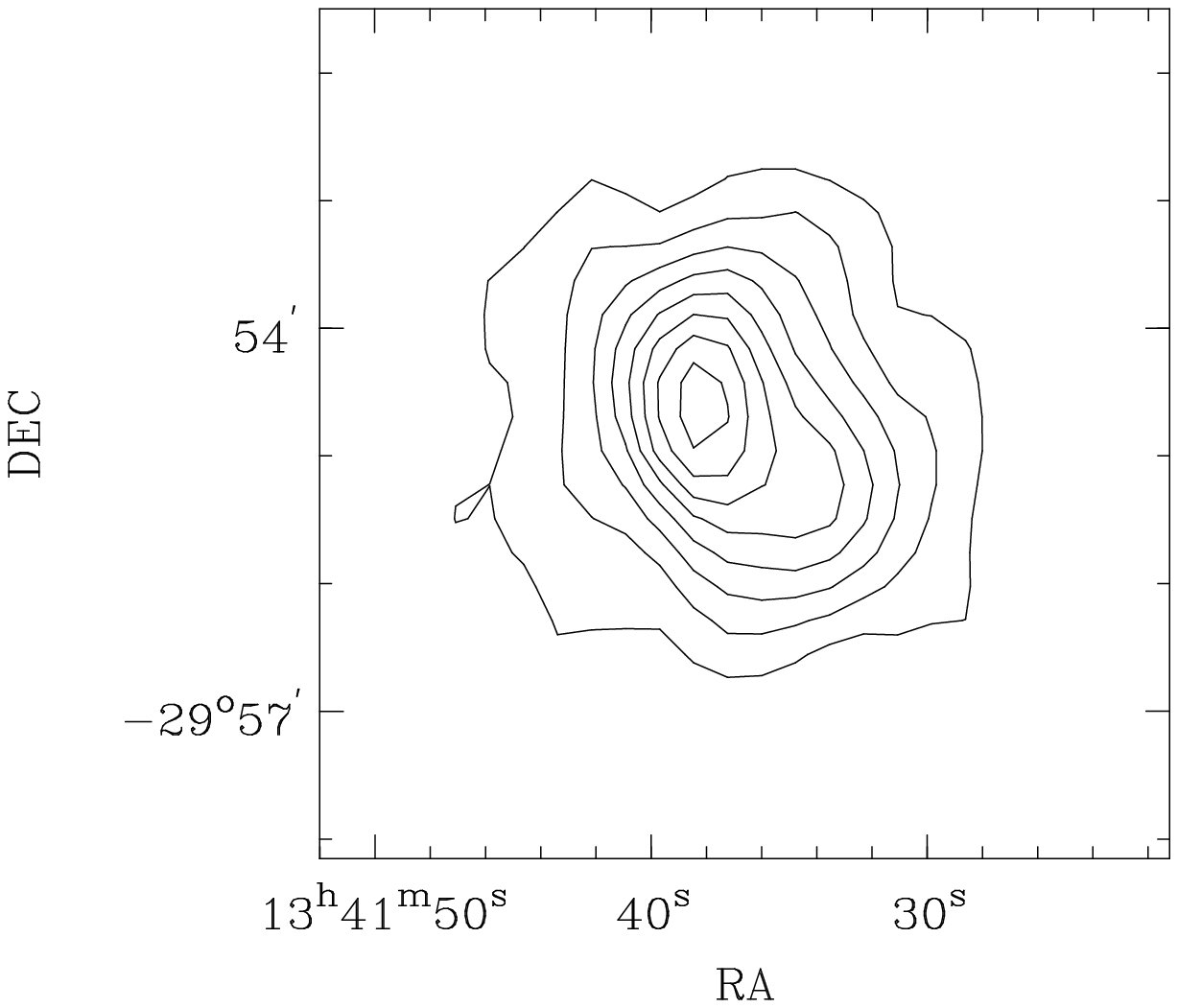}

\epsfxsize=0.33\hsize
\epsfbox{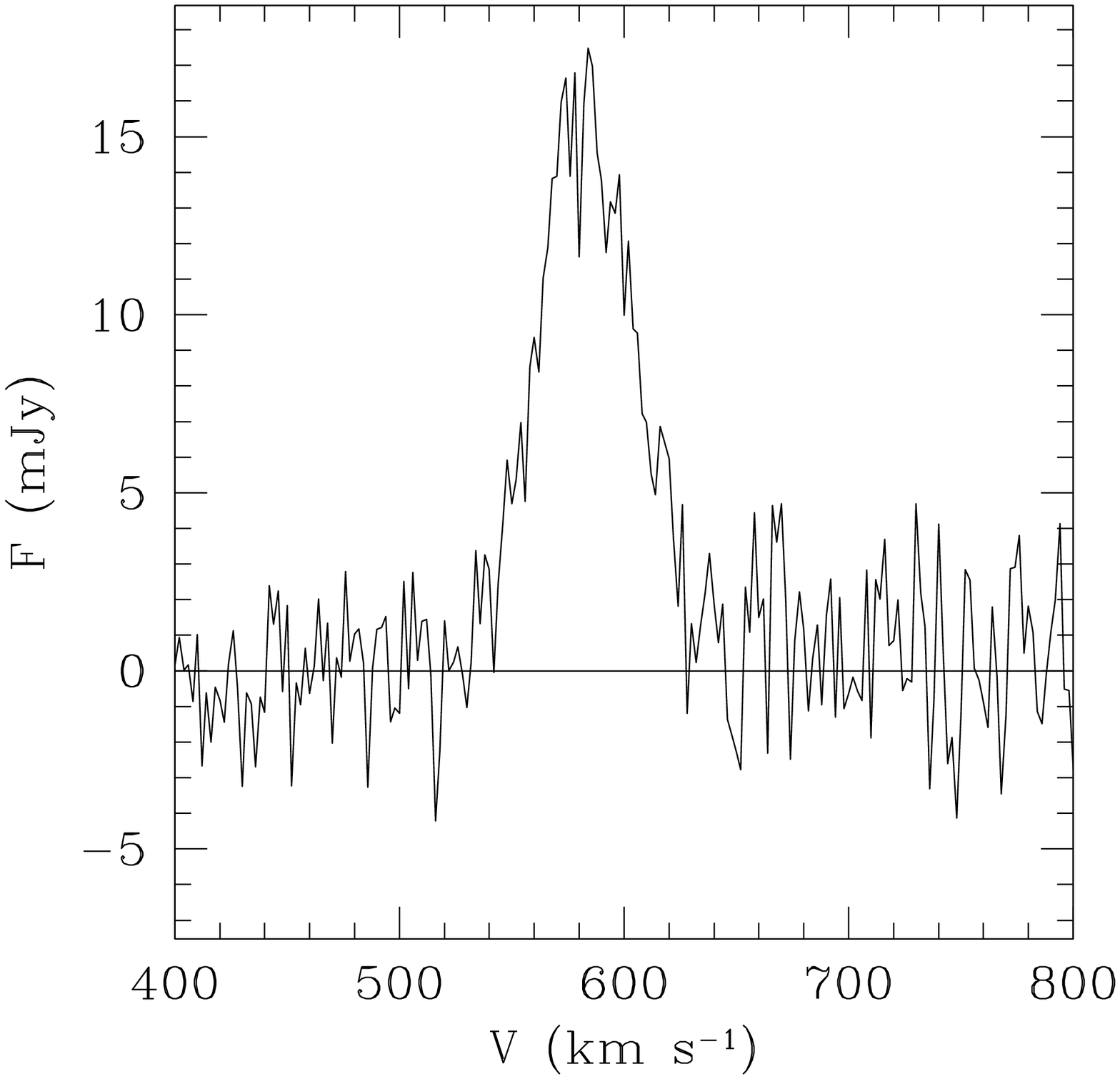}
\epsfxsize=0.33\hsize
\epsfbox{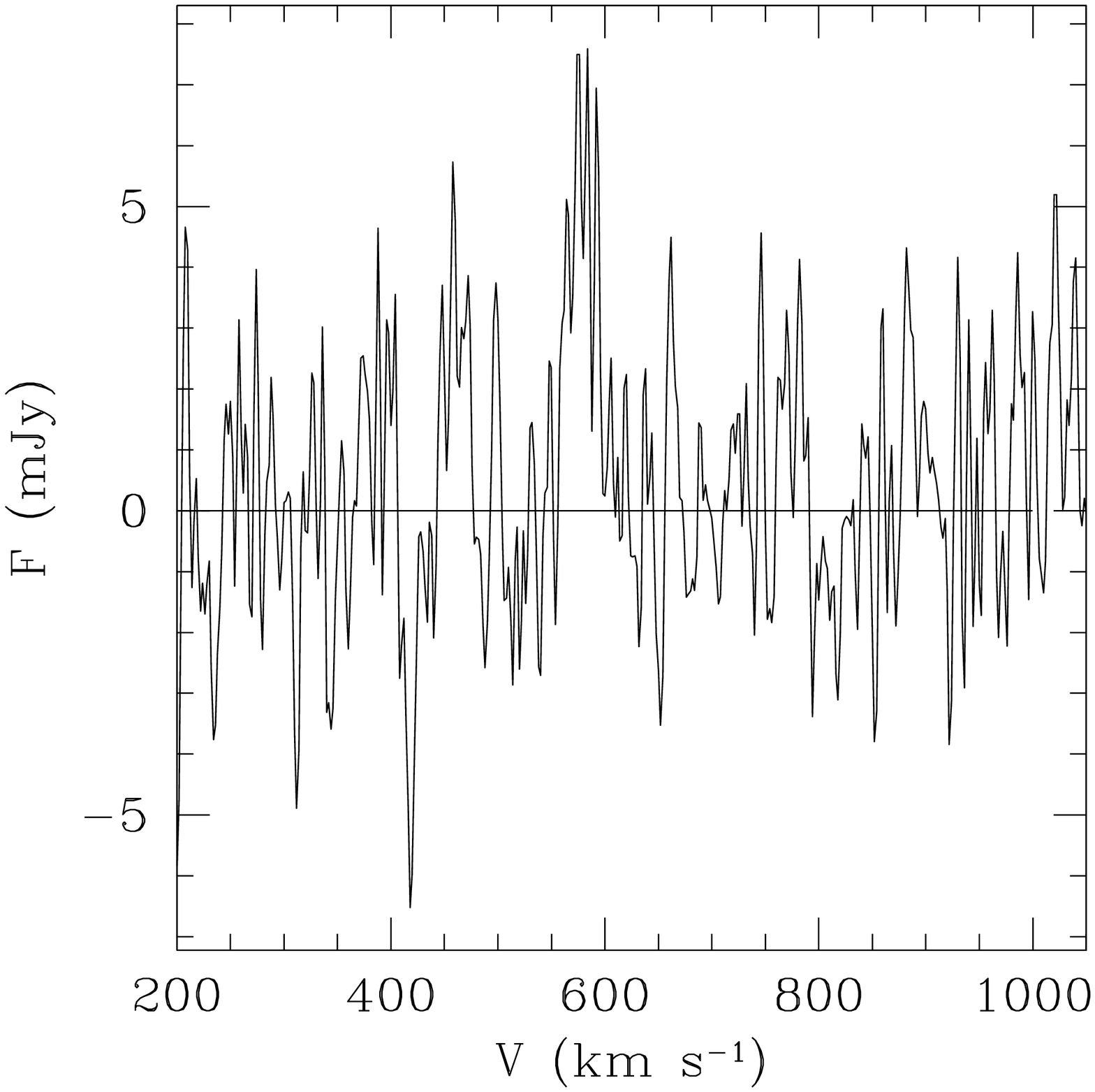}
\epsfxsize=0.33\hsize
\epsfbox{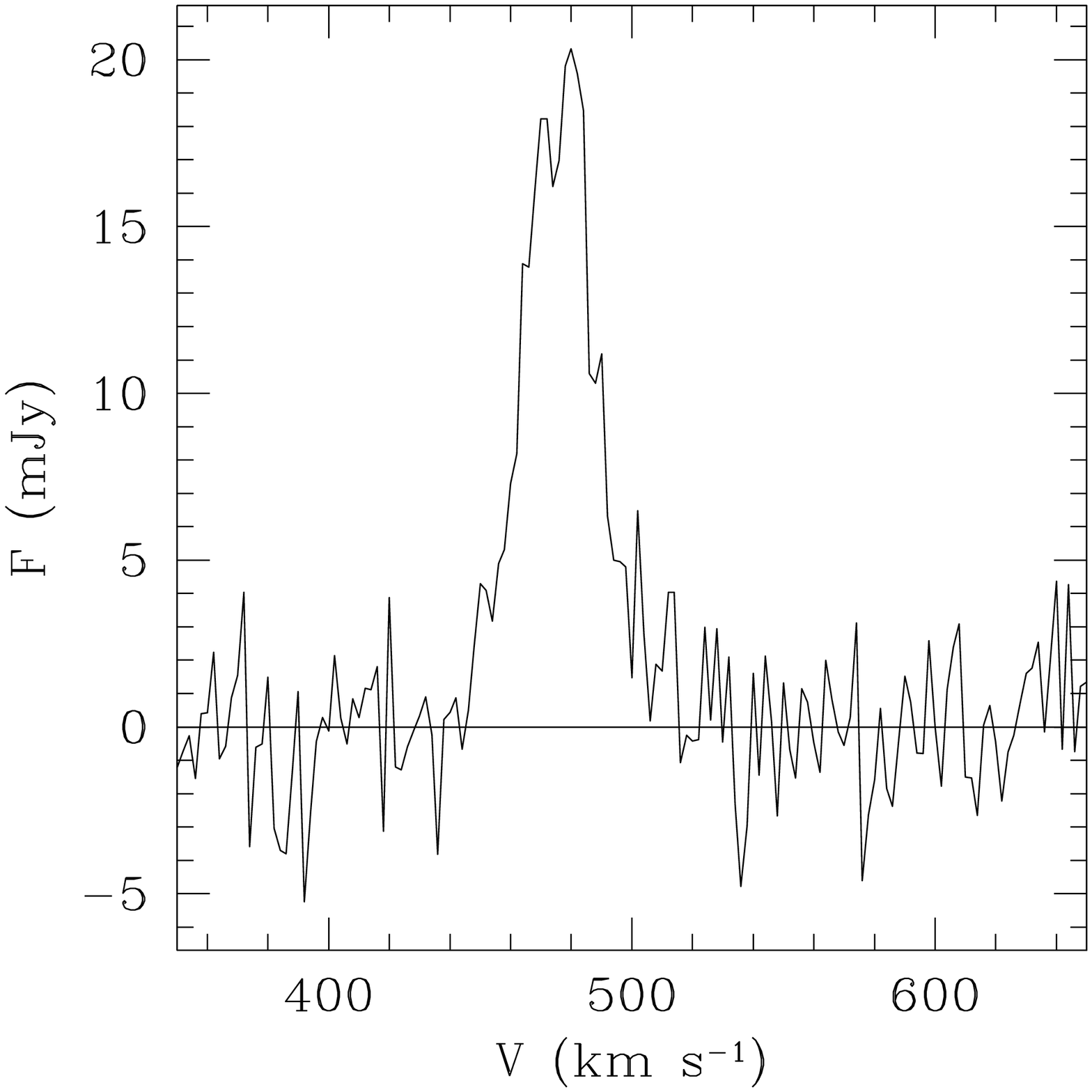}
\caption{Integrated HI column density maps and integrated
spectra of the detected known galaxies (M83 is not displayed here).
Left: ESO 444-G084. Center: IC 4316. Right: N5264.
Top row shows zeroth-moment maps. Contour levels are: [ESO 44-G084, left]: (1, 3, 5, 7, 9, 11)$\times 10^{20}$ cm$^{-2}$; [IC 4316, center]: (0.5, 1, 1.5, 2)$\times 10^{20}$ cm$^{-2}$; [N5264, right]: (1, 2, 3, 4, 5, 6, 7, 8)$\times 10^{20}$ cm$^{-2}$. Bottom row shows the integrated spectra. The spectrum of IC 4316 (center) has been hanning-smoothed. }
\end{figure*}

\subsubsection{Blind search}

\paragraph{Method}

The cubes were searched using an automatic search program in order to
avoid any human biasses.  Our proceedure extracts a spectrum at each
spatial pixel and hanning smoothes it. A heavily smoothed (75 pixel
boxcar) version of the spectrum is then subtracted to remove baseline
ripple and other large-scale variations.  This inevitably loses very
wide profiles, but these observations are tuned to look for narrow
signals. The difference spectrum is then searched for significant
signals at different velocity resolutions. The spectra were searched
at their original resolution and after smoothing with 3, 5, 9, and 15
pixel (6, 10, 18, 30 km s$^{-1}$) boxcars (one velocity pixel is 2 km
s$^{-1}$). The first 30 and the last 50 channels were not searched as
bandpass effects cause the noise there to be higher. The effective
search range is therefore 192 - 1600 km s$^{-1}$. Signal due to the
known galaxies was also removed prior to the automated search.

We decided to place our cut-off at the $5\sigma$ level. At this
significance level the number of false detections expected due to
Gaussian statistics is still managable: in each cube we expect to find
$\sim 13$ noise pixels with values $>5\sigma$ (and a similar number
with values $< -5\sigma$). At the $4\sigma$ level we would expect
$\sim 1400$ false noise signals per cube, which is clearly unfeasible.

The actual detection process is a two-step one. Due to the large
amount of data  which must be inspected at
four velocity resolutions, the computation speed is I/O limited. Our
algorithm reads the datacube only once, one spectrum at a time, but
the proceedure both (1) keeps a table of all signals with an absolute
value $>4.5\sigma$ at any resolution evaluated relative to the noise
level measured for each spectrum, and (2) accumulates the 
necessary statistical information to compute the noise level for each
frequency plane in the cube. Each velocity range (frequency channel)
might be corrupted by narrow RFI that would raise the noise level in
certain individual planes in the cube.  This permits a second pass
through the list of tentative $>4.5\sigma$ detections to determine
whether they are significant relative to the noise level in the
velocity place where they were identified.  Only the signals that made
the $5\sigma$ cut at one or more velocity resolutions on the second
pass were retained.  This process was applied at all velocity
resolutions with only one pass through the entire dataset.

To analyze the data at lower spatial resolution, we also applied the
full detection algorithm to a new set of spatially smoothed cubes.

The output is a set of lists of positions with absolute flux values
$>5\sigma$ for each velocity resolution.  Obviously, there will be
many cases where a signal is detected at multiple velocity
resolutions.  We have therefore also produced a list where all double
or multiple detections (defined as $|\Delta R| < 2$ pixels [32$''$]
and $|\Delta V| < 3$ pixels [6 km s$^{-1}$]) have been removed.  We
will refer to these as ``double detections'' or ``doubles''.

For comparison, with this method IC 4316 is detected at the
8.0$\sigma$ level at the 15 channel (30 \kms) resolution, at the
$7.1\sigma$ level for 9 channels (18 \kms), at $6.3\sigma$ for 5
channels (10 \kms), at $5.6\sigma$ for 3 channels (6 \kms) and at
5.0$\sigma$ at full spectral resolution (2 \kms). It should be kept in
mind that IC 4316 is located $17'$ away from the pointing center. Had
it actually been observed in the center of the field, then all
significances would have been twice as high.

\paragraph{Results}

In the absence of real signal and the presence of Gaussian noise we
expect each cube to contain an equal number of ``detections'' with
values above $5\sigma$ and below $-5\sigma$.  For theoretical Gaussian
noise there ought to be $\sim 26$ such pixels (13 positive and 13
negative) in each full-resolution cube, although counting statistics
and the fact that we expect real data to have a slightly non-Gaussian
noise distribution, means that the actual number of detections may vary
a little.

Fig~4 plots the number of positive detections in each cube versus the
number of negative ones. In the absence of any real signal we expect
an equal number of cubes to have a positive and a negative excess.
This is clearly the case for the Cen data (within the uncertainties
mentioned before).

Table~3 summarizes the results. There is no clear evidence that the
Cen data cubes contain anything more than noise at the $>5\sigma$
level.

\begin{figure}
\epsfxsize=\hsize
\epsfbox{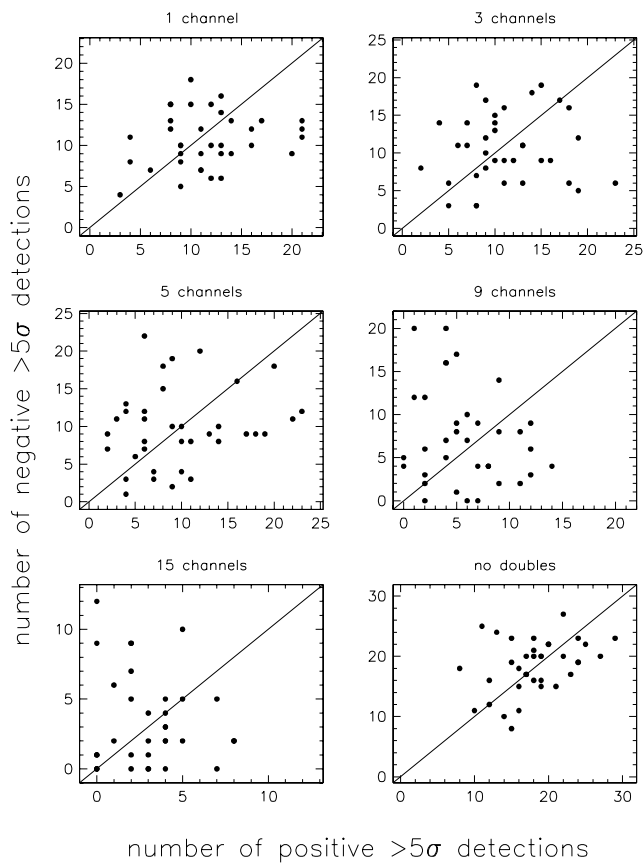}
\caption{Number of positive $>5\sigma$ detections versus
number of negative $<-5\sigma$ detections in the Centaurus data.  Each
point represents one data cube. Different panels show different
velocity resolution.  Panel at the bottom-right combines all velocity
resolutions but double detections have been removed.}
\end{figure}

\begin{table}
\caption{Positive and Negative Detections}
\begin{tabular}{rccc}
\hline
           & \multicolumn{3}{c}{Number of cubes with} \\
Resolution & pos $>$ neg & pos $=$ neg & pos $<$ neg \\
\hline
\multicolumn{4}{c}{Centaurus}\\
\hline
1 channel  & 19&	1&	16\\
3 channels &19&	1&	16\\
5 channels &18&	2&	16\\
9 channels &16&	1&	19\\
15 channels&17&	5 &	14\\
no doubles &16&	3&	17\\
\hline
\multicolumn{4}{c}{Sculptor}\\
\hline
1 channel  &7&	2&	13\\
3 channels &8&	2&	12\\
5 channels &7&	4&	11\\
9 channels &6&	8&	8\\
15 channels&1&	18&	3\\
no doubles &6&	3&	13\\
\hline

\end{tabular}
\end{table}

\begin{figure}
\epsfxsize=\hsize
\epsfbox{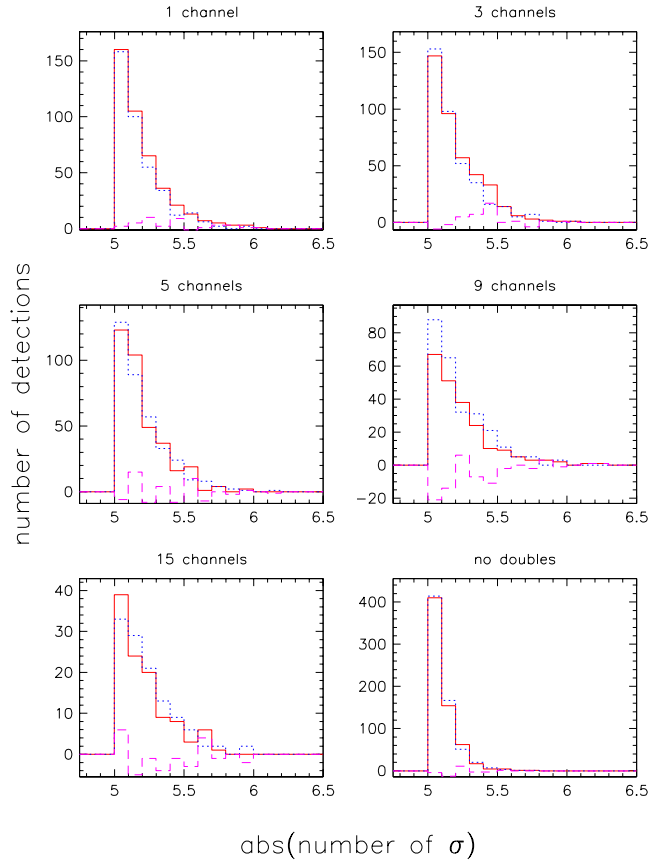}
\caption{The distribution of the number of detections $> 5\sigma$ versus
the absolute values of their significance in the Centaurus data. The
full lines represent the positive detections, the dotted lines the
negative detections and the dashed lines the difference
histogram. Different panels show different velocity resolution.  Panel
at the bottom-right combines all velocity resolutions but double
detections have been removed.}

\end{figure}

A different way of looking at the results is to compare histograms of
the peak flux distributions of the ``detections.'' This is done in
Fig.~5, for each velocity resolution separately, and for the case
where double detections have been removed.  The positive and negative
distributions are within their Poisson-errors indistinghuishable.
Again we see no evidence for the presence of real objects in the data.

\subsubsection{Spatial Smoothing}

The beam size of $65''\times43''$ is optimized to detect objects of
$\sim 1$ kpc size at the distance of the Cen A group. In an attempt to
find more extended objects we have smoothed the data by a factor of 5
spatially, resulting in a beam size of $325''\times 215''$ (this is
roughly the maximum scale structure that the interferometer
configuration that we have used is able to detect). We performed a
similar blind search on the spatially smoothed data, but found no new
$>5\sigma$ objects. The noise in the smoothed maps is 30 mJy, which
for a channel separation of 2 \kms, translates in a $5\sigma$ column
density of $4.8 \cdot 10^{18}$ cm$^{-2}$. After smoothing to 10 \kms,
the column density limit is $1.1 \cdot 10^{19}$ cm$^{-2}$.

\subsection{Sculptor}

\subsubsection{Known objects}

In contrast with the Cen survey, there were no known extragalactic
objects present in the survey area. We did detect an extended HVC
centered at $00^h25^m24^s, -31^{\circ}37^m57^s$, $v=-70$ km
s$^{-1}$. This HVC is prominently visible in the HIPASS data. In most
cubes a bright very narrow linewidth feature ($\sim 5$ km s$^{-1}$),
presumably Galactic emission, is present at $v=+4$ km s$^{-1}$.  These
features were subsequently blanked in the data in order to facilitate
the blind search.

\subsubsection{Blind search}

The Scl data cubes were searched in the same way as the Cen cubes.
Fig.~6 shows for each velocity resolution separately the number of $>
5\sigma$ detections plotted agains the number of $<-5\sigma$
detections.  Table~3 summarizes these results. As with Cen there is no
evidence for an excess of positive detections. There is a hint of an
excess of negative detections, however this is at most a $\sim
1.5\sigma$ difference. The excess negative detections occur in a
narrow velocity range adjoining the (blanked out) Galactic emission,
and are therefore most likely artefacts caused by the relatively
strong Milky Way emission.  Figure~7 shows the noise histograms for
both positive and negative signals. As with Centaurus we find no
evidence for the presence of anything other than noise.

\begin{figure}
\epsfxsize=\hsize
\epsfbox{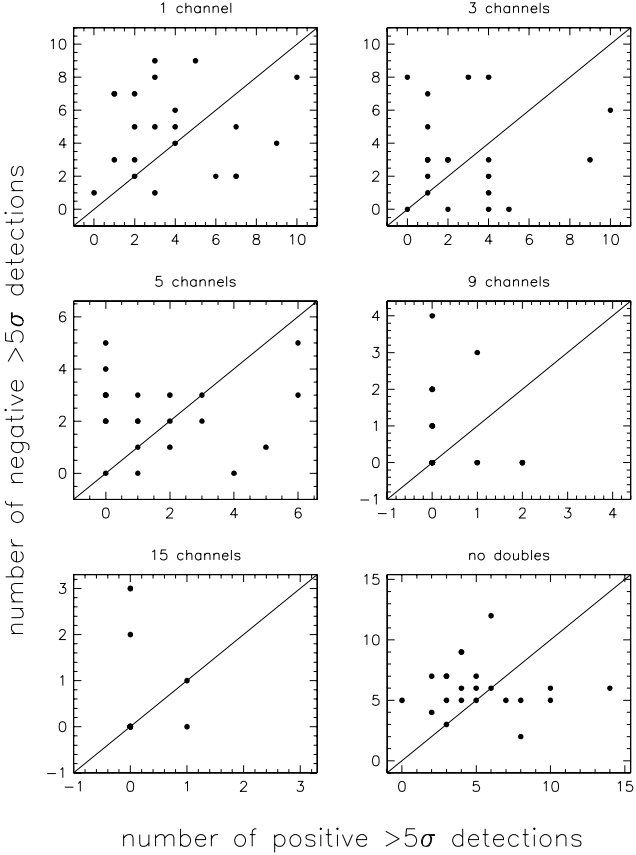}
\caption{Number of positive $>5\sigma$ detections versus number of negative $<-5\sigma$ detections in the Sculptor data.
Each point represents one data cube. Different panels show different
velocity resolution.  Panel at the bottom-right combines all velocity
resolutions but double detections (at one or more smoothings) have
been removed.}
\end{figure}

\begin{figure}
\epsfxsize=\hsize
\epsfbox{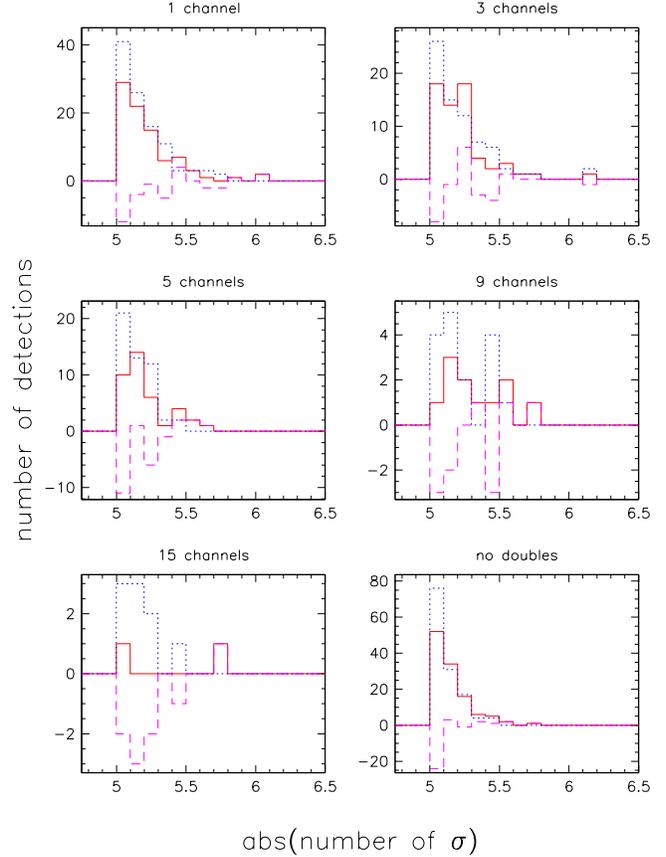}
\caption{The distribution of the number of detections $> 5\sigma$ versus the
absolute values of their significance in the Sculptor data. The full
lines represent the positive detections, the dotted lines the negative
detections and the dashed lines the difference histogram. Different
panels show different velocity resolution.  Panel at the bottom-right
combines all velocity resolutions but double detections (at one or
more velocity resolutions) have been removed.}

\end{figure}

\subsubsection{Spatial Smoothing}

We smoothed the data to one-fifth of its original spatial resolution
(resulting beam $285'' \times 145''$). Apart from the aforementioned
HVC and the Galactic signal we did not detect any additional
detections with a significance $>5\sigma$. For pointings A1-D4 (E1-F4)
the noise was 47 (61) mJy, which for a channel spacing of 2 \kms,
translates in a $5\sigma$ column density sensitivity of $1.3\ (1.6)
\cdot 10^{19}$ cm$^{-2}$, and after smoothing to 10 \kms to a
column density sensitivity of $2.8\ (3.7) \cdot 10^{19}$ cm$^{-2}$.

\section{Discussion}

\begin{figure}
\epsfxsize=\hsize
\epsfbox[10 140 600 718]{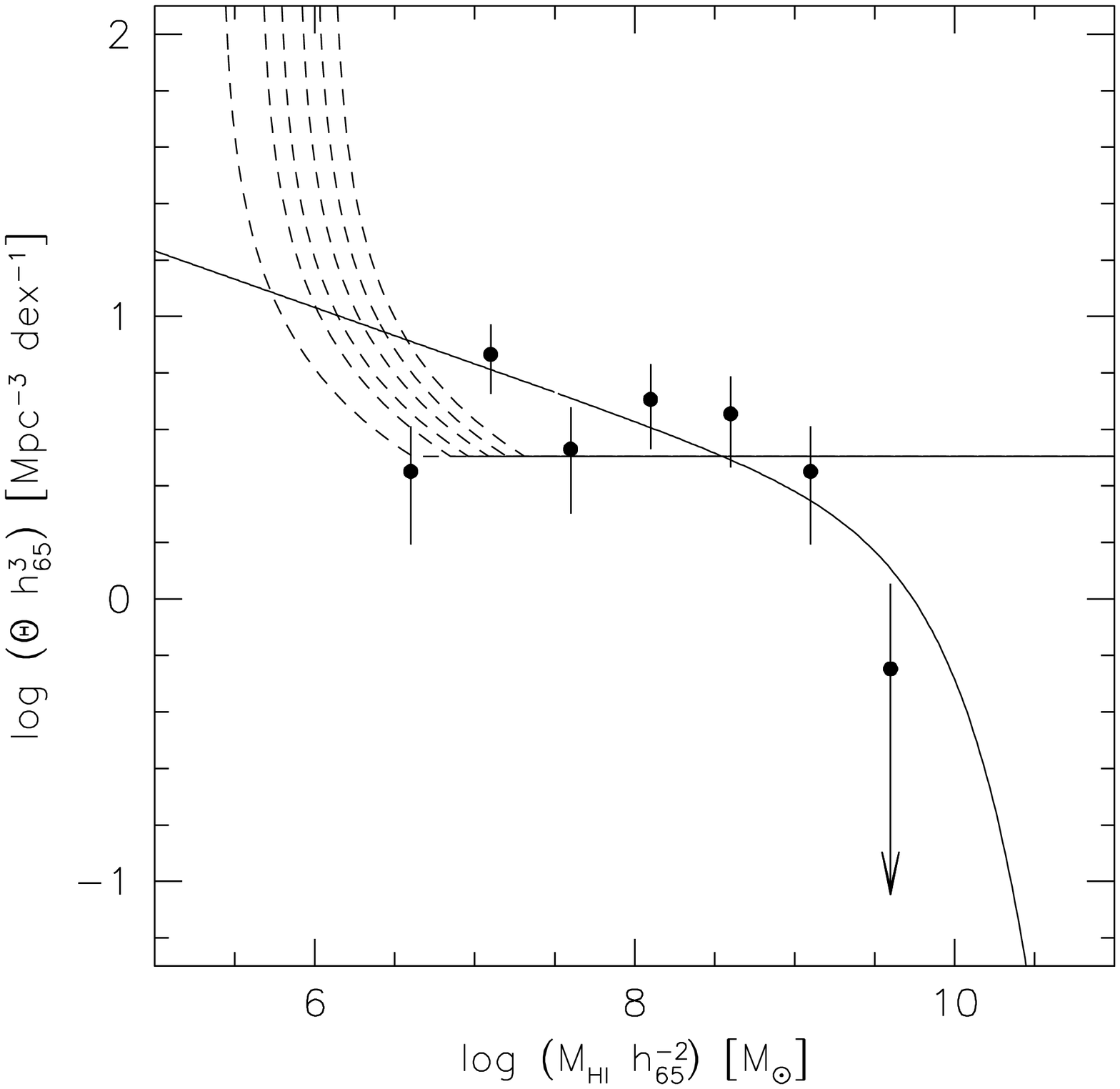}
 \caption{Space densities of objects as a function of \hi\ mass in the
 Centaurus and Sculptor groups.  The solid curve indicates an \himf\
 with a faint-end slope of $\alpha=-1.2$ (Zwaan et al.  1997), scaled
 vertically by a factor 300 to compensate for the overdensity in the
 groups.  The dashed lines represent the reciprocals of the surveyed
 volumes within the group, for objects with velocity widths of 2, 6,
 10, 18, and 30 \kms\ (from left to right).  Detections are expected
 where the \himf\ exceeds these lines.  Since no new objects were
 detected in this survey, steep faint-end slopes of the \himf\ can be
 excluded.}  
\end{figure}

Fig.~8 shows the space density of galaxies in the two surveyed groups,
in combination with the sensitivity of the surveys.  The solid dots
represent the space density of known group members with measured \hi\
masses \citep{cote}, which are calculated by dividing the number of
galaxies per 0.5 dex bin, by the total volume of the group.  For the
group radii we adopt the values $R=0.64$ Mpc for Cen A
\citep{vandenbergh} and $R=0.66$ Mpc for Sculptor \citep{freemanhi}.

We estimate a mean overdensity of the groups by vertically scaling the
Schechter fit to the field \hi\ mass function from \citet{zwaan97} so
as to fit the points in Fig.~8.  Using this method we arrive at an
overdensity in the two groups of 300 times the cosmic mean.  This
number should be regarded as a rough indication of the groups'
overdensity because the calculation is heavily dependent on the
definition of the groups' boundaries.  The calculation assumes that
the groups are spherical while the actual shape of the groups are more
complicated. This is especially true for Cen A as can be seen from
Fig.~1, but possibly also for Sculptor \citep{jerjen}. Furthermore, due
to substructure in the groups, the actual overdensity is a strong
function of the position in the groups.  For the conclusions reached
in this paper, the value of the overdensity is however not crucial.

The dashed lines in Fig.~8 show the reciprocal of the surveyed volumes
for different assumed velocity widths of \hi\ clouds or galaxies.
They are calculated for optimal spectral smoothing for velocity widths
of 2, 6, 10, 18, and 30 \kms\ (from left to right).  The shape of the
curves is determined by the (approximately) Gaussian primary beam of
the survey.  Larger volumes are searched for higher mass galaxies
since these can be detected out to larger distances from the centre of
the primary beam.  The curves level off at \hi\ masses of
approximately $10^8 M_\odot$, because we assume that no detections can
be made at separations larger that 35 arcmin (equal to the FWHM) from
the centre of the primary beam.  

From the information presented in Fig.~8 it is now trivial to
calculate how many detection the survey should have turned up assuming
different shapes of the \hi\ mass function.  This number can be
evaluated by taking the integral over \hi\ mass function multiplied by
the survey volume.  Assuming the \himf\ as presented in Fig.~8, and a
velocity width of 6 \kms, the survey should have detected $2.6$
galaxies.  This number drops to 1.8 and 1.7 for velocity widths of 30
and 50 \kms.  More detection are expected if the \himf\ would show an
upturn below $M_{\rm HI}=10^{7.5} M_\odot$, as has been proposed by
\citet{schneider}.  If the faint-end of the HIMF would have a steep
slope with $\alpha=-1.5$ below $M_{\rm HI}=10^{7.5} M_\odot$, the
expected number of detections would rise to 5.3, 2.4 and 2.1 using
velocity widths of 6, 30 and 50 \kms.  Clearly, even steeper
extensions of the \himf\ are excluded by our present survey.

The null-result of this group survey can be used to place tight
constraints on the space density of small \hi\ clouds in galaxy
groups. Fig.~9 shows the combined constraint on the \hi\ mass of
intragroup clouds and the space density of clouds.  The lines show the
95\% confidence levels on the existence of cloud populations.  These
lines are calculated assuming Poisson statistics and the fact that no
new clouds were detected. Regions of parameter space above each line
are excluded by our survey.

\begin{figure}
\epsfxsize=\hsize
\epsfbox[10 140 600 718]{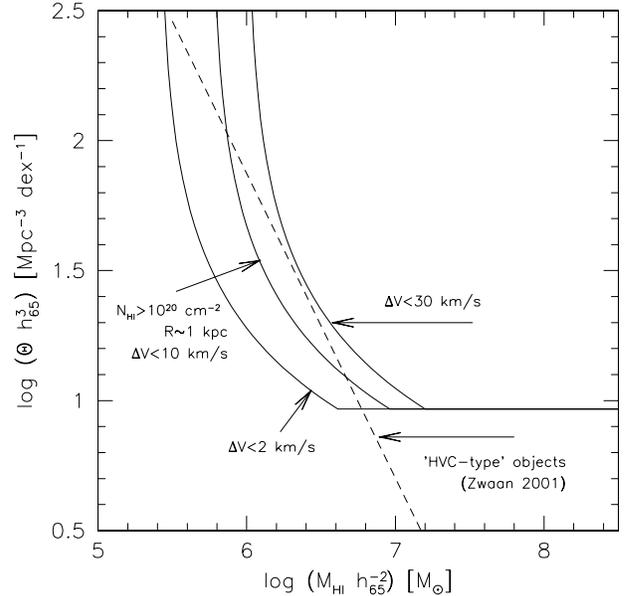}
 \caption{95\% confidence upper limits to the space density of \hi\
clouds for objects with velocity widths of 2, 10, and 30 \kms.
Also shown as a dashed curve is the upper limit to the space density of
\hi\ clouds derived by Zwaan (2001) by means of Arecibo survey in five
galaxy groups. The limits set by the present survey are stricter, but
only apply to clouds with column densities $N_{\rm HI}>10^{20} \rm cm^{-2}$. }

 \end{figure}

For comparison, we also show the 95\% confidence upper limits on cloud
populations as calculated by \citet{zwaan01} by means of a pointed
Arecibo survey in five galaxy groups.
 The limits on intragroup \hi\
clouds from the present survey are slightly stricter than those from
\citet{zwaan01}, but care should be taken with the interpretation of the
results.  The surface brightness sensitivity of the Arecibo group
survey was much better, resulting in very low detectable \hi\ column
densities of $1.0 \times 10^{18}\rm cm^{-2}$ ($5\sigma$) per 16 \kms\
for gas filling the beam.  The full resolution \hi\ column density
sensitivity of the present ATCA survey is $1.0 \times 10^{20}\rm
cm^{-2}$ ($5\sigma$) per 10 \kms.  The two surveys are therefore
sensitive to different populations of \hi\ clouds, the Arecibo survey
to lower column densities, the ATCA survey to lower \hi\ masses.

The beam area of the smoothed cubes is $\sim 25$ times larger than in
the original resolution cubes, and the column density sensitivity is a
factor of 10 better, so in principle the smoothed cubes could be used
to constrain the density of more diffuse \hi\ clouds, and would be
suitable to assess the problem of extragalactic HVCs
\citep{blitz}.
However, as our column density sensitivity and survey volume are less
than those of the surveys by \citet{zwaanbriggs00} and
\citet{zwaan01}, our limits are not as strong as the ones
presented in those surveys.  

Using the parameters of compact HVCs from Putman et al.  (2002), we
expect them to have sizes of approximately $2.5'$ at the distance of
the surveyed groups, and having peak column densities of $1.4 \times
10^{19}\rm cm^{-2}$ over 35 \kms.  This means that most of the CHVCs
in the groups would escape detection in the present survey.

In summary, the current survey supports the
conclusion reached previously that HVCs are not objects with
\hi\ masses $\sim 10^7 M_\odot$ distributed throughout galaxy groups,
but does not improve on them.

\section{Conclusions}

One of the results of the CDM theory of formation of structure in the
universe is that every galaxy group should be filled with a few
hundred small mass halos.  Optical surveys for these objects have only
turned up $\sim$ 10\% of the predicted numbers.  If the density of the
primordial \hi\ in the satellites is sufficiently high to prevent
ionization by the meta-galactic uv-background, they should be
detectable in 21cm \hi\ surveys. This motivated us to perform a
sensitive 21cm search in two nearby galaxy groups: the Centaurus A and
Sculptor groups.

We have used the Australian Telescope Compact Array to observe 36
fields in the Cen A group in the velocity range 192--1600 km s$^{-1}$
and 22 fields in the Sculptor region in the velocity range $-300$-500
km s$^{-1}$. The surveys are sensitive to \hi\ masses down to a few
times $10^{6}$ \Msun, which is an improvement of a factor 10 on
previous surveys. These surveys were tuned to pick up narrow profiles
that might in single-dish surveys with a coarser velocity resolution
have been confused with RFI.

We find no new detections, though observations of known galaxies show
that we do indeed reach our theoretical mass limit. This null-result
rules out faint \himf\ slopes steeper than $\alpha=-1.5$ below
$10^{7.5}\ M_{\odot}$ in these groups, and puts tight constraints on
the occurence of compact \hi\ clouds in other groups.

We have also searched the data for extended and diffuse \hi\ clouds,
and our null-detections support the results derived by e.g.\
\citet{zwaanbriggs00} and \citet{zwaan01}  that 
HVCs are not distributed in the intra-group environment on scales of 1
Mpc.

\end{document}